\documentclass[aps,rmp,twocolumn,groupedaddress,showkeys,altaffilletter,nofootinbib,a4paper,floatfix]{revtex4}

\usepackage{dcolumn}
\usepackage{bm}

\ifx\pdfoutput\undefined 
\usepackage[dvips]{graphicx}
\usepackage{color}
\else
\usepackage[pdftex]{graphicx}
\RequirePackage[colorlinks,hyperindex,plainpages=false,citecolor=red]{hyperref}
\fi
\begin{document}


\title{\textit{\scriptsize\mdseries\setlength{\baselineskip}{3ex}\mbox{European Space Agency Symposium
``Ensuring Long-Term Preservation and Adding Value to Scientific and Technical Data'', 5~-~7 October 2004,
Frascati, Italy}\\\vspace{3mm}} Providing Authentic Long-term Archival Access to Complex Relational Data}


\author{Stephan Heuscher\,$^a$}
\author{Stephan J\"armann\,$^b$}
\author{Peter Keller-Marxer\,$^c$}
\author{Frank M\"ohle\,$^d$}
\affiliation{\mbox{\normalsize Swiss Federal Archives, Archivstrasse 24, CH-3003 Bern, Switzerland}
\\
\small{\\ $^a$stephan@heuscher.ch, $^b$stephan.jaermann@bar.admin.ch,
$^c$peter.keller@acm.org, $^d$frank.moehle@bar.admin.ch}}



\date{July 19, 2004}


\begin{abstract}
\parbox{15cm}{We discuss long-term preservation of and access to relational databases. The
focus is on national archives and science data archives which have to ingest and integrate data
from a broad spectrum of vendor-specific relational database management systems (RDBMS).
Furthermore, we present our solution SIARD which analyzes and extracts data and data logic from
almost any RDBMS. It enables, to a reasonable level of authenticity, complete detachment of
databases from their vendor-specific environment. The user can add archival descriptive metadata
according to a customizable schema. A SIARD database archive integrates data, data logic, technical
metadata, and archival descriptive information in one archival information package, independent of
any specific software and hardware, based upon plain text files and the standardized languages SQL
and XML. For usage purposes, a SIARD archive can be reloaded into any current or future RDBMS which
supports standard SQL. In addition, SIARD contains a client that enables `on demand' reload of
archives into a target RDBMS, and multi-user remote access for querying and browsing the data
together with its technical and descriptive metadata in one graphical user interface.}
\end{abstract}

%
\maketitle

The urgency of deep-infrastructure solutions for long-term digital preservation and archiving was
clearly formulated in the late 90's of the past century by national archives and libraries
\cite{TaskForceReport,napUniverse1,Universe2,Roth95,Roth99,interpares,oldBiblio} as well as space
agencies and institutions in earth observation, oceanography, and astronomy
\cite{DADworkshop,OAISgeneral}. The finding of problem statements and strategies is still in
progress and was recently supported by a charta of the United Nations \cite{UNESCOcharter}. Deep
and lasting solutions are still not available, and for national archives and libraries (faced with
very heterogeneous types of content) it is broadly accepted that digital collections are growing at a
rate that outpaces their ability to manage and preserve them \cite{NSFworkshop}.

There is an ongoing process of recognition that long-term digital preservation poses similar
problems in diverse disciplines. Despite of different vocabularies used by different communities,
research and development in the field can only be successful in a joint effort. During the past
decade, however, decisive progress was achieved in analytical and conceptual work
\cite{StateOfArt,CedarsGuide,napERA}. Furthermore, the Open Archival Information System (OAIS)
reference model became widely accepted in diverse disciplines and covers a full range of archival
information preservation functions including ingest, archival storage, data management, access, and
dissemination. It has become the international standard ISO 14721:2003 \cite{ISO-14721-OAIS} and
may be applicable to any archive as it does not refer to specific implementations or archival
strategies but merely provides a common terminology and functional framework to discuss 
different implementation approaches.

Many current development projects focus on the archiving of digital images (digitized photographs
or paper documents) and sound recordings, while more complex types of digital information have been
neglected, even though their relevance for governmental and scientific activities has drastically
increased during the past decade. In particular, a recent international workshop on the long-term
preservation of databases \cite{ErpanetWorshopDatabases} revealed that many archives do have a
long-standing practice and experience in ingesting and preserving relational data, but their daily
work is constrained to the treatment of rather simply structured data sets, requires extensive
manual work by archives personnel, and does not allow for smooth and standardized integration of
data and descriptive meta data on a level required to ingest, preserve, and provide access to
complex relational databases.

In this paper we present a method and application named ``Software-Invariant Archiving of
Relational Databases'' (SIARD), developed at the Swiss Federal Archives. It completely detaches
typed relational data from almost any relational database management system, while still retaining
most of the original data logic and integrating data and metadata in one archival information
package that is based on text files and standardized technologies. In Section
\ref{ComplexityRelevance} we discuss the technical and intellectual complexity of relational data
in modern database systems, the resulting problems for long-term preservation, and its relevance to
archives. The objectives in the development of SIARD are described in Section \ref{Objectives},
while Section \ref{WorkflowFeatures} covers SIARD's system architecture, workflow, features, and
development platform.

\section{\label{Introdcution}Introduction}

Relational data is one of the oldest forms of structured information representation, intuitively
used already centuries before the ``digital age''. With the rise of computer technology, the
introduction of mathematical formulations of the relational data model in the mid-20th century, and
the international standardization of a corresponding data definition and query language, relational
data has become an omnipresent method to organize data for electronic data processing in almost
every field of work, form business activities to scientific research and government administration.

During the past two decades, usage has developed from processing single-table data files with
specific application software to generic relational database management systems (RDBMS). These have
internal mechanisms for logical and physical organization of arbitrary relational data models, are
able to physically store terabytes of data, cover rich data types (including internal procedural
code), enable multi-user transactions, and provide internal data life-cycle management. The
definition, representation, management, and query of relational data was thereby standardized and
separated from specific application logic and application software that operates on the database.
As a result, RDBMS have become core components of almost any type of digital information system.

It is obvious that this development has decisive impacts on the work of those institutions which
are charged to collect or accept digital data from various data sources, to make it broadly
accessible, and to preserve it over decades: national archives and libraries, science data
archives, or business companies being under special legal regulations for long-term data retention
(like, for example, the pharmaceutical sector \cite{erpaPharma}).

\section{\label{ComplexityRelevance}Complexity and Relevance}
\subsection{\label{TechnicalComplexity}Technical Complexity of Relational Data}
One consequence of the progress in database technology is that relational data and relational
databases become highly complex. They often consist of hundreds of linked tables (i.e. physical
representations of relational entities\footnote{We will not discuss the relational model in this
article \cite{MeltonSQL0,Codd}. It will be sufficient to think of \textit{tables} which consist of
one or more table \textit{columns} and one or more table \textit{rows}. The points of intersection
of columns and rows contain \textit{data items} which have a \textit{data value} and a \textit{data
type}. If (and only if) there never occur duplicate rows in a table, then the table is a
\textit{relation}, and its columns and rows are also called \textit{attributes} and \textit{tuples}
(or \textit{records}), respectively. Using relational algebra and calculus, several relational
tables can be managed and manipulated jointly.}), which makes it impossible to handle and query
table data outside an RDBMS if these links become broken or cannot be managed automatically
anymore.

Furthermore, any data item in a RDBMS has a precisely defined data type and domain. (Entire tables
or integral parts of them may likewise have types.) Apart from basic data types (for example
integer and real numbers, dates, and character strings), low-level types like large binary objects,
complex and inheritable user-defined types, and multi-lingual character encodings are widely used
in modern databases. As for the linking of tables, the connection between the data and its data
types and domain definitions is not preserved when table data is trivially exported to external
plain text files.

In addition to those entities and features mentioned above, modern RDBMS include, for example:
\begin{itemize}
    \item Check constraints and assertions: For a single column of a table or a set of
    entire tables, assure that changing or entering data does never violate defined data types,
    quantitative restrictions, or value domains. In particular, it can be assured that data items
    in a table column will never be empty.
    \item Views: Assemble selected parts of several tables and operate on them as customized, virtual
    tables.
    \item Triggers: Force the RDBMS to initiate timed operations on data when user-specified
    conditions are met, for example log and audit user activities.
    \item Functions (basic and user-defined): Perform numerical calculations, conversions, or
    character operations on data items or sets.
    \item Stored Procedures: Store and execute programs inside the RDBMS to perform common
    or critical tasks which are not part of the specific application software outside the RDBMS.
    \item Foreign Keys: Ensure referential data integrity, i.e., automatically prevent that values can
    be stored in rows of one table if there are no corresponding values in referencing entries within
    the database.
    \item Grants and Roles: Define user profiles and assign or withdraw privileges,
    for example to create new tables or access certain parts of the database.
\end{itemize}

For long-term preservation in national archives, relational data is collected from many different
database systems and has to be retained and kept processable and accessible for decades. It is
therefore essential to store and maintain the databases independent of any specific and short-lived
products (or at least transferring them all into only one preferred product). In fact, most RDBMS
use the same language for the definition of the internal logical organization of data, namely the
declarative (i.e. non-procedural) Structured Query Language (SQL). Despite of its name, the scope
of SQL also includes the definition of data structure and the manipulative operations on data
stored in that structure \cite{MeltonSQL0}.

The development of SQL started in the 1970's, leading to the international standard ISO/IEC 9075 in
1987, and evolved in four main stages through SQL-89, SQL-92, SQL:1999, and recently SQL:2003
\cite{ISO-9075-SQL}, while the size of the standard has grown from 120 to over 2'000 pages.
SQL:2003 and SQL:1999 are fully upward compatible with SQL-92. The standard language keywords are
structured in three subsets: Data Definition (DDL), Data Manipulation (DML), and Data Control and
user authorization (DCL). To increase acceptance by vendors, the standard defines three levels of
conformance and implementation: entry, intermediate, and full level. The mandatory part of SQL:1999
and later is called the ``Core'' of SQL and described in Part 2 (Foundation) and Part 11 (Schemata)
of the standard.

Aside from revisions to all parts of SQL:1999 (e.g. new data types and functions that return entire
tables \cite{MeltonSQL1}), the 2003 edition contains the new part: ``SQL/XML'' defines a minimal
handling and integration of text-based data structured by the Extensible Markup Language XML
\cite{XMLspec}. This includes \cite{MeltonSQL2} mappings between tables and XML documents, SQL data
types and XML Schema \cite{XMLSchemaSpec} data types, and RDBMS implementation-specific character
sets to Unicode \cite{UnicodeSpec,ISO-10646-UCS,UTFspec} character encodings. Additionally, the
related standard ``SQL Multimedia and Application Packages ''ISO/IEC 13249:2003 \cite{ISO-13249}
defines a number of packages of generic data types common to various kinds of data used in
multimedia and application areas, to enable storage and manipulation of such data in a relational
database.

\subsubsection{\label{SQLstandardized}Standardized -- Really?} SQL is an internationally
standardized and comprehensive language for the definition, description, query, and manipulation of
relational data and databases. It is widely used since almost 25 years, developed upward
compatible, and will probably play a key role for another 25 years. Since most RDBMS are based on
SQL (and most vendors claim compliance with the standard) one could assume that relational database
definitions are independent of any specific RDBMS product. Unfortunately, this is far from being
true. In contrast to standardized programming languages like ISO-C or ANSI-Fortran, SQL-based
database layouts and SQL code can rarely be ported between different RDBMS without major
modifications and loss of functionality.

There are two main reasons for severe incompatibilities. First of all: Although the SQL standard
today comprises over 2'000 pages, it is far from being fully self-contained. In contrast, SQL:1999
explicitly identifies 381 so-called implementation-defined items and 137 so-called
implementation-dependent items \cite{MeltonSQL0}. Their implementation is left open for any
manufacturer of RDBMS products. As long as a manufacturer completely documents all
implementation-defined items, the product can rightly claim to comply with the SQL standard, though
it differs from all competing products. (The precision of the SQL integer data type is a simple
example of an implementation-defined item.)

The second reason is that most of today's RDBMS implement only (and sometimes faultily) the core
and the entry level of the standard completely, but add plenty of non-standard, product-specific
enhancements, leading to different ``flavors of SQL''. These include \cite{SQLdifferences} new
additions to or modifications of, for example, data types, functions, operators, behavior and
syntax of SQL statements. Additionally, almost all RDBMS products use their own procedural
programming languages for stored routines (Oracle PL/SQL, Postgres PL/pqSQL, Microsoft T-SQL,
PL/Perl etc.) rather than implementing the standard's procedural language SQL/PSM.

Finally, it is an almost trivial remark that modern RDBMS move the physical storage of the
data from the operating system (file level) to the application level (the internal storage of the
RDBMS).

\subsubsection{\label{SQLarchiving}``SQL for Archiving''?}

Considering the imponderables discussed in the previous subsection we can draw the following
conclusion with respect to archival institutions: If they have to collect and ingest relational
data from various database management products for the purpose of integrating them, to preserve them
over long periods of time, and to make them broadly accessible, then these institutions are faced
with a Sisyphean task: The data they have to preserve for long is locked up in short-lived
obsolescent and complex software products from a vast diversity of manufacturers, making the
longevity of the data heavily depend on the availability of the products and their versions, the
support by vendors, and the existence of the manufacturers. One lesson learned is that long-term
preservation of relational databases is much more than just making backups of export or dump
files from database management applications, though IT professionals usually use 'archiving' and
'backup' as synonyms.

It is a tempting idea to demand a ``SQL for Archiving'', defined as a subset of the ISO-9075-SQL
standard. It would leave database designers the choice to restrict their databases to layouts
suitable for long-term preservation (and transfer of databases among competitive RDBMS products, of
course). A similar effort is undertaken by government agencies and industry representatives to
define an ISO standard "PDF for Archiving, PDF/A" \cite{PDFA} based on the popular Portable
Document Format (PDF) specification 1.4 \cite{PDFspec} by Adobe Systems Inc. and the ISO 15930
standard PDF/X \cite{PDFX}. However, the current ISO-9075-SQL standard is probably not suited for a
similar  attempt: too many important items are left open as implementation-defined and would
require adding deep descriptions of implementation details in the standard.

It should be mentioned that since its first edition, ISO-9075 requires SQL implementations to
provide a feature called ``SQL-Flagger" (feature F812 in SQL:1999) which is able to identify and
signal certain kinds of non-standard SQL language extensions used in the specific implementation.
The feature is implementation-dependent and its intention is to assist SQL programmers in producing
SQL language that is portable among different conforming SQL-implementations. Standard SQL flagging
is only required for the entry level of SQL. In fact, most major RDBMS manufacturers have
implemented SQL-Flaggers for SQL-92 in their products. The reason is that the U.S. Government
implicitly requires conformance with the entry level of SQL-92 for all SQL products in federal
procurements. (Conformance with higher levels may be specified explicitly.) The minimum
requirements for conformance with entry-level SQL-92 as well as specific features of the
SQL-Flagger are specified in the Federal Information Processing Standard (FIPS) 127
\cite{FIPS-127}.

The purpose of FIPS SQL is ``to promote portability and interoperability of database application
programs, to facilitate maintenance of database systems among heterogeneous data processing
environments, and to allow for the efficient exchange of programmers among different data
management projects'' \cite{FIPS-127}. Although not explicitly mentioned in FIPS 127, its intention
also supports long-term preservation of databases. The U.S. National Institute of Standards and
Testing (NIST) used to validate SQL implementations to conform with FIPS-127. Although this
procurement specification is still in force nowadays, it was not updated to the 1999 and 2003
editions of SQL, and NIST ceased its product validation in 1997. (A suite of automated validation
tests for SQL-92 implementations is still freely available from NIST \cite{FIPS-127}.) To our
knowledge, none of the major RDBMS products include SQL-Flaggers for SQL:1999 or SQL:2003 as it
would be required by the respective standard editions. Our application SIARD (described in Section
\ref{WorkflowFeatures}) has its own, built-in SQL validator. Aside from this, we are aware of only
one (third party and commercial) tool which provides such functionality \cite{MimerVal}.

We conclude from our work, that standard (i.e. generic) SQL may be reasonably exploited for
long-term preservation purposes only when data and data logic is actively extracted from database
management systems by specialized ingest tools which map different "SQL flavors" to generic SQL,
and transparently trace and document those parts which cannot be mapped.

\subsection{\label{Intellectual}Intellectual Complexity and Access}

Aside from the technical aspects discussed so far, a successful long-term preservation of databases
is only possible if the intelligibility and comprehensibility of the database and its data can be
preserved as well. To keep data understandable and meaningful it is indispensable to collect enough
technical as well as non-technical metadata and handle it as an integral part of a database
archive. Otherwise, there will be a rare chance to understand the meaning and value of the
database's content decades after it was archived.

However, most of the meta-data necessary to enable long-term \textit{intellectual} access to the
data is not deposited in the database but is provided to the archives on separate (and often
paper-bound and hard to grasp) documents. On the one hand, this includes precise and complete data
dictionaries, code lists, narrative descriptions of the naming, meaning and usage of single
database objects. On the other hand, it is descriptive and archival metadata about the context,
creation, purpose, usage, or chain of custody of the original database and the RDBMS used to
operate it.

The problem of descriptive metadata that supports intellectual accessibility has to be considered
in a broader context of trusted digital repositories \cite{Trust}: Any serious long-term
preservation strategy for any kind of digital content aims to guarantee continuous

\begin{itemize}

\item \textbf{Integrity}: protection of the data from unintended and intended harm;

\item \textbf{Intelligibility}: understandability and comprehensibility of the data;

\item \textbf{Authenticity}: authentication (of authorship and provenance) and reliability;
(of evidence)

\item \textbf{Originality}: data structure and functionality ``as close to the original as
possible'';

\item \textbf{Accessibility}: technical readability and usability.

\end{itemize}

However, due to overall technical obsolescence in a digital environment, these are competitive and
conflicting goals. Each archival institution will therefore have to establish its own measures and
priorities among the five above-mentioned criteria. The measures will be primarily ruled by
metadata, and many authors address the interplay between obsolescent technical infrastructures and
continuing guidance of metadata \cite{Trust,interpares,Stephens,Dollar}.

Moreover, since long-term preservation is a costly and laborious task, effective appraisal methods
are increasingly important instruments. The value of information and evidence contained in
databases is often determinable only through the purpose, design, and context of usage of the
original database application. Again, these criteria are measured by means of metadata provided by
the data producers.

Another level of intellectual complexity is introduced by an increasing amount of interlinked,
federated and temporal database systems which makes it difficult to determine the correct spatial
and temporal scope for extracting data for long-term preservation: A database that is selected for
archiving may refer to time-dependent master data in another database, or a database does not
overwrite or delete any outdated data but rather records all data modifications, using timestamps
as multiple primary key components (e.g. valid-time state tables, tracking logs, backlogs, etc.).
It will be a challenge for archives to keep such spatial and chronological dependencies and
interrelations understandable and traceable, particularly across sequential accessions from the
same database.

\subsection{\label{CurrentPractices}Current practices}

Relational data kept by national archives usually consists of plain text files that contain
tabulated data with fixed-length or delimited columns \cite{ErpanetWorshopDatabases}. Very often,
these files are only derivatives of the original database, produced by denormalization of the
original data model to reduce the number of single tables. The data files are usually accompanied
by paper documents or microfiches that provide data dictionaries and other descriptive information
necessary to understand the content, provenance, and context of the data. At the time when the data
was originally transferred to the archives, it has been scrutinized to reveal any inconsistency
with the paper-bound documentation.

In typical archival environments, the descriptive documents are kept separately in cardboard boxes
which are stored in air-conditioned shelve vaults, while the electronic data files are stored on
labeled magnetic tapes or cartridges. Tapes and cartridges are recopied every 5 to 10 years to
prevent data loss from degradation of the magnetic media by physical and chemical processes.

One reason for the deficiency in providing long-term physical and intellectual access to data from
archived databases is a high heterogeneity of data formats and the fragmentation of data sets into
isolated data files, which are hard to handle in bulk. But primarily it is the habitual bipartite
treatment of digital data in most archives: While the data itself is accessible and processable by
electronic systems, the metadata necessary to understand the data's content, provenance, and
context, is bound to paper documents and often incomplete, erroneous, outdated, not standardized,
and hard to grasp. That's why even a small data collection requires cumbersome and expensive manual
work to overcome this divide between electronic and paper documents.

It can be easily concluded from the discussion of the previous sections that the situation will
become even more critical in the future: The advance in information technology puts more and more
obstacles on the path that digital information has to pass on its way from producers to the
archives. Proprietary, closed data formats and technologies quickly become obsolete, and
heterogeneous and complex data structures further impede comprehensive data integration in the
archives.

Today's archives become aware that digital information in their custody becomes increasingly
volatile rather than persistent. This is particularly true for relational databases since the
development of new preservation techniques has mainly focused on other forms of digital content
such as images or sound recordings which are easier to handle and more attractive to a broad
audience. In contrast, methods for the long-term preservation of data from modern database
management systems with increasing complexity and data in the order of terabytes has been
neglected.

\subsection{\label{NationalArchives}Relevance to National Archives}

Relational data is probably the oldest and most widespread type of information among the electronic
digital holdings of national archives, typically dating back to the 70's and 80's of the past
century. (In fact, even the oldest collections are of no age compared to usual archival time
scales.) These collections comprise data from almost any field of activity of governmental
institutions and are thus of high information and evidential value, and are increasingly important
to researchers in diverse fields such as History, Sociology, Politics, Economics, Meteorology, or
Geography.

Nevertheless, national archives nowadays are hardly ever able to provide broad access to their
digital data sets on a level of usage comparable to paper-bound or analog electronic holdings. (A
few archives have put a lot of effort into it, though, and provide public access to selected
collections through the world-wide web \cite{DataOnline}.) Most national archives are still far
from enabling the public to locate government information regardless of format.

In terms of archival appraisal, databases may serve various purposes which are not in the primary
focus of national archives to preserve evidence in business records of the federal administration.
(Business records are evidence of what an agency has done or decided.) In the opinion of many
archivists, databases are mainly used by the administration to store and manage registers, master
data records, and statistical (e.g. census) data, thus may be at best considered as finding aids.
This may be true for databases of the past.

Nowadays, modern database systems are widely used for managing and recording business processes and
transactions. Thus they have become integral parts of almost any record and document management
system or E-government web site, and thus often contain high evidential value. It has become an
essential necessity for archives to assess the evidential value of databases kept in RDBMS, and to
develop appropriate criteria and guidelines for such assessments \cite{Zurcher,VERS}.

Criteria and guidelines are also required for appropriate archival description of relational
databases since they are rarely integrated into any filing plan of the agency. In consequence,
there is only a purely technical ordering of database records which will not correspond to the
logical ordering of business records (e.g. belonging to a business case, dossier, or document). A
business record may comprise several technical table records and may extend across several tables,
and a single database table is often not a meaningful entity for archival description. It can be a
challenge to identify and precisely describe something like a dossier, business record, or document
in a relational database made up of dozens of tables \cite{Zurcher,Shepherd}.

\subsection{\label{Research}Relevance to Natural Science Data and Research}

It is broadly accepted that experimental scientific data has often a high long-term value
\cite{napUniverse1,Universe2,napEarthandSpace,napLongterm,napRoleOfScienceData}, and the urgency of
solutions for their long-term preservation has been pointed out many times during the past decade
\cite{NAA,CODATA}. By contrast with national archives, science data archives focus on the
information value, while evidential value of scientific data is rarely considered. Moreover, there
are completely different criteria and schemes for appraisal and description of scientific data
\cite{ErpanetWorshopScience,PVToulouse}.

Scientific data may have a high long-term value because, for example, it cannot be reproduced (e.g.
climate and oceanographic data) or it was produced at enormous costs (e.g. high-energy physics
experiments or space flight missions). As theories evolve and new questions arise, archived data
may be reconsidered and re-evaluated in future research, and may turn out to be of essential
scientific value. For example, satellite data from the 60's and 70's of the past century turned out
to be essential for current research on global warming. Long-term preservation solutions of
scientific and technical data is also needed, for example, in the automobile and aeroplane industry
since construction data has to be retained for cases of insurance claims.

However, the driving force in development of solutions for long-term preservation of scientific
data is to provide sustained global data access and exchange between globally distributed research
collaborations \cite{PVToulouse,napBitsOfPower,Caltec}.

\section{\label{Objectives}Objectives}

From the discussion of the previous sections we briefly formulate objectives for research and
development in the field of long-term preservation of relational databases. A solution for the
archiving of relational databases

\begin{itemize}
\item has to enable, to a good level of authenticity, permanent retention of the original
internal data structure, the referential data integrity, and all technical and non-technical
meta-information needed to keep the data technically accessible and intellectually understandable
over the long term;

\item must ensure that the data remains utilizable and processable by future data processing systems;

\item must be able to completely detach databases from its proprietary
database management software, hardware, and operating system environments;

\item must completely rely on widely accepted and internationally standardized technologies;

\item shall no longer require any specific software, maintenance, or
administration for at least ten years, and reasonably longer if the full documentation of applied
standard technologies remains available, provided that the physical bit-streams of all files remain
intact;

\item shall easily enable the reload to current and future relational database management
system products, and thereby allow queries in almost the same complexity (on the database level) as
in the original system from which the database was archived from;

\item shall support the acquisition and standardization of non-technical data (which is usually
not available from the physical database) across different business units and persons, and seamless
integration with data and technical metadata.

\end{itemize}

\section{\label{WorkflowFeatures}Solutions: SIARD Workflow and Features}

The method and Java application SIARD (``Software-Invariant Archiving of Relational Databases'')
meets the aims outlined in Section \ref{Objectives}. The development of SIARD was a subproject of
the strategic project ARELDA (Archiving of Electronic digital Data) of the Swiss Federal Archives
and the Swiss Federal Administration.

SIARD consists of three stand-alone software applications, named A0, A1, and A2. Each application
may be used by different people at different places during different stages of the workflow:
database administrators and application responsibles of the RDBMS, records managers of the data
producers, IT professionals and general personnel of the archives, or archives customers. The
output of each application is used as the input of the next stage's application. Unfinished work
can be saved any time and resumed later. All three applications are independent of the computer
platform (cf. subsection \ref{Development}) and location. Communication between SIARD components
and the RDBMS can take place either locally or through a TCP/IP network (which will be the usual
case). All applications have XML-based, external configuration files that may also be edited
manually.

\begin{figure}
\includegraphics[height=222mm,keepaspectratio]{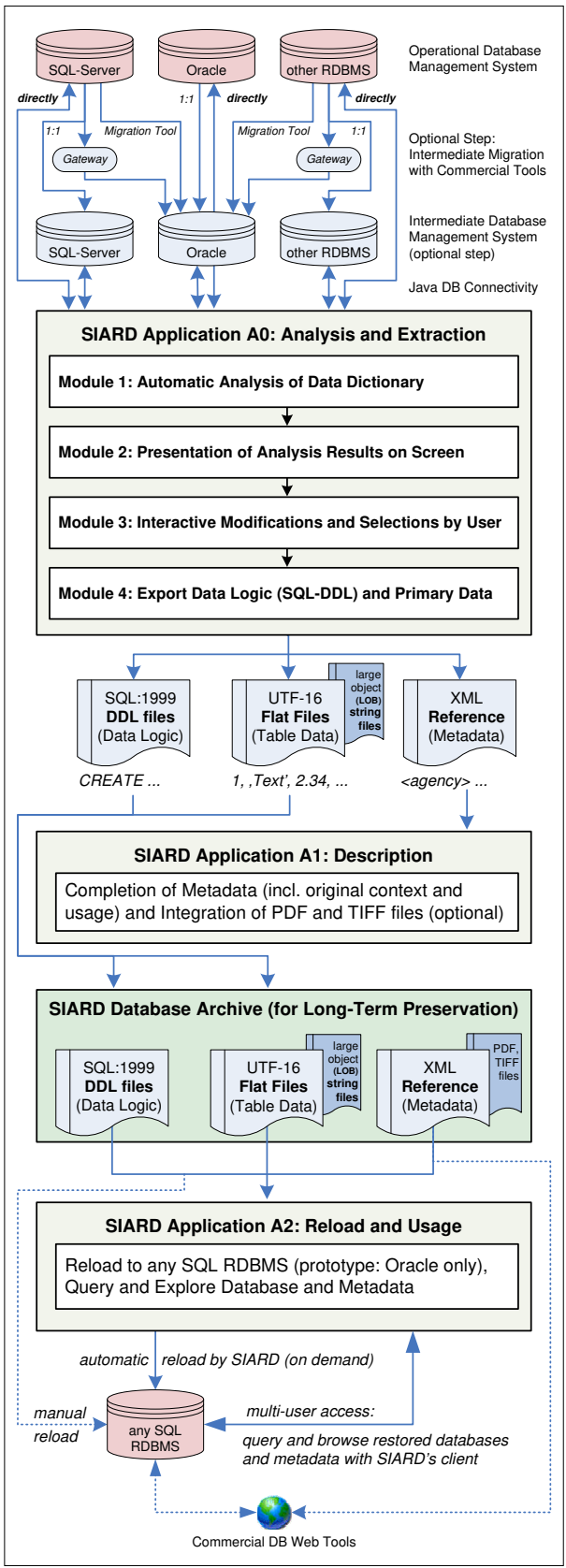}
\caption{\label{FigWorkflow}Overview of SIARD components and workflow: The relational database to
be archived is extracted from a RDBMS (top), the SIARD archive is created for long-term
preservation (middle) and finally restored in a RDBMS (bottom) for usage.}
\end{figure}

Figure \ref{FigWorkflow} summarizes the SIARD components and workflow. The RDBMS which contains the
database to be archived is at the top of the figure. As examples, Figure \ref{FigWorkflow} shows
two commercial RDBMS products (Microsoft SQL-Server and Oracle) as well as an unspecific ``other
RDBMS'' which could be an Open Source product, for example. (SIARD may also be used with Microsoft
Access, that works best when Access is used in the SQL-Server compatibility mode.) Figure
\ref{FigWorkflow} sketches various set-ups for starting work with SIARD:

\begin{itemize}
\item Direct connection to the operational RDBMS.
\item Migration of the database from the operational RDBMS to another RDBMS (including another
instance of the same RDBMS product) using migration tools that are supplied by the vendors. (We
denoted this intermediate RDBMS with ``Oracle'', but it could be any other product.)
\item Connection to the operational RDBMS through a vendor-supplied transparent gateway.
\end{itemize}

The first method (direct connection) is the usual case. However, direct connections of SIARD to the
operational RDBMS may not be allowed or desired, for example due to security reasons or because the
RDBMS is a high-availability system. (However, we emphasize that SIARD does \textit{not} alter the
RDBMS in any way! It solely performs read-only access operations on the RDBMS.)

But there are other possible reasons to add this intermediate step: For example, using transparent
gateways or migration of the database from the original to another RDBMS may help to get more out
of SIARD since it provides so-called ``expert modes'' for some specific products. These expert
modes take into account and exploit product-specific features. Thus it may be advantageous to
previously migrate the database from the original RDBMS to a RDBMS product for which SIARD provides
an expert mode. Another reason could be that pre-processing of the database is required prior to
archiving, for example data conversions or filtering. (SIARD can only exclude entire columns or
tables from archiving.) Of course, such operations will never be performed on a production database.

As shown in Figure \ref{FigWorkflow}, the first SIARD application A0 analyzes and extracts a
database from the RDBMS (with guidance by the user), and creates the archive files. After these
tasks have been completed, the XML reference file (which contains all meta information about the
results of the extraction) is loaded into application A1. In A1, the user adds further, mandatory
and optional non-technical metadata on all levels of the object hierarchy (database, tables,
columns etc.) as well as context metadata defined by the archival institution. Being complemented
this way, the XML reference file is written back to the SIARD archive which is now ready for
long-term preservation. The third application A2 comes into play when the archived database is
requested for access and usage by a costumer or required for other dissemination purposes. The
reload is initiated through A2 by either the customer itself or by archive staff. We will describe
the features of the three SIARD applications in more detail below.

Several tests have proven the applicability of the solution. The complexity of the tests performed
so far range from tens of tables and a few thousand rows up to 250 tables and 250'000 rows from
several commercial and open source database management products. The main difficulties were, as
expected, proprietary, non-standard extensions to SQL in all RDBMS products,
implementation-specific character sets, and extraction of some specific metadata from the system
dictionaries, respectively.

\subsection{\label{A0}A0: Analysis and Extraction}

This is undoubtedly the core of SIARD since it determines the quality and extent of the database
archive as far as technical aspects are concerned. A0 usually will be operated by the RDBMS
application responsible (who has knowledge of the databases content), maybe assisted by a database
system administrator (DBA) who has a deeper understanding of the technical background of the
specific RDBMS. (Of course, A0 can also be used within the archival institution to migrate older
non-SIARD data collections, for example.)

There are two ways to connect to the RDBMS with A0. The straight forward method is to allow A0 to
connect as a DBA. Although A0 solely performs read-only operations on the database, this method is
is not recommended since it requires disclosure of the DBA password. The recommended method is that
a DBA creates a new user in the RDBMS (named SIARD-A0, for example) and grants to it only those
rights which are required by A0 to be operated properly. This also allows fine-tuning of privileges
to an extent where A0 can only see exactly those parts of the database that are in fact subject to
archiving \footnote{Privilege tuning for A0 is described in the user manual and may depend on the
specific RDBMS product used.}.

At startup, A0 asks for the Java Database Connectivity (JDBC) \cite{Java} parameters to be used.
The user either must enter the parameters manually into a panel, or else open up an XML-based A0
configuration file. (Usually, a DBA will provide a configuration file.) Afterwards, A0 asks for the
access mode to be used. There are so-called ``expert modes'' for specific RDBMS products as well as
a generic mode, used if no specific expert mode is provided by SIARD or the type of RDBMS is
unknown. At the moment, there are only two expert modes (Oracle 7/8/9 and Microsoft SQL-Server
7/2000). Expert modes provide a broader range of database objects and metadata that can be archived
since they exploit product-specific features. The generic mode has a rather narrow focus and solely
uses the standard functionality of JDBC (which is continuously evolving from version to version,
though.) New expert modes may be easily programmed and added to A0 without changes in existing
code.

\begin{figure*}
\includegraphics[width=179mm,keepaspectratio]{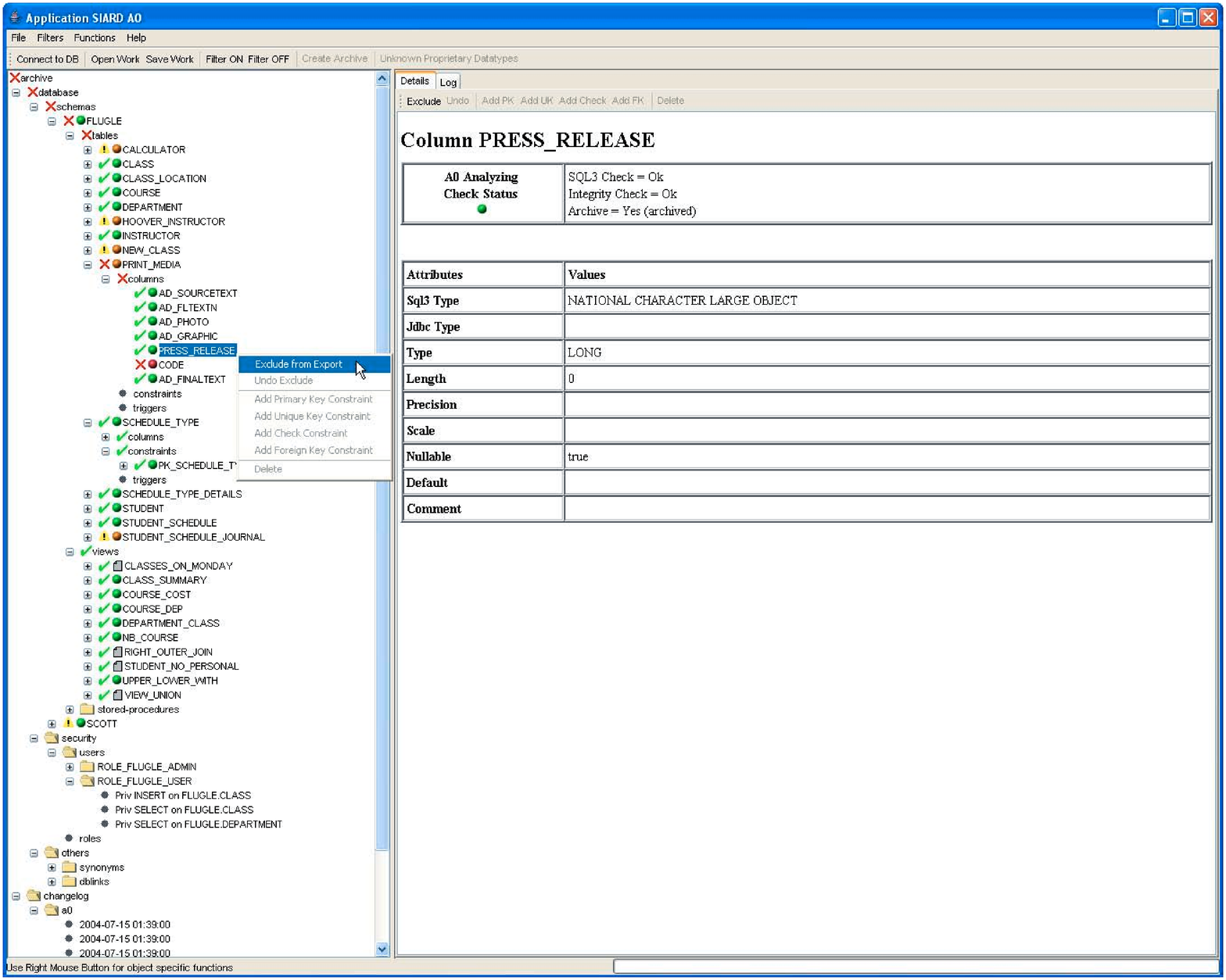}
\caption{\label{FigA0}A0's workbench after database schemata (FLUGLE and SCOTT in this example)
were analyzed. The right-hand pane shows the technical metadata and analysis results for the
database object that is selected from the database object tree in the left-hand pane. Colored
bullets and symbols indicate where user decisions are required to proceed.}
\end{figure*}

\subsubsection{\label{a0-analysis}Automatic Analysis and Mapping}

After successful connection to the RDBMS, A0 lists all\footnote{To be more precise: All schemata
are listed only if A0 has logged in with DBA permissions. Otherwise, if A0 is logged in as an
individual SIARD user, the list of schemata depends on the privileges that were granted by the DBA.
Schemata and tables which are solely for internal purposes of the RDBMS are automatically ignored
by the expert modes.} schemata\footnote{If the RDBMS supports catalogs, e.g. Microsoft SQL-Server,
these are listed before the schemata.} in the database and asks the user to select all or only some
of them to be archived by SIARD. Afterwards, the database schemata (data dictionaries) are analyzed
by A0's built-in SQL parser and validator. Additionally, A0 checks for data integrity and
identifies, for example, isolated tables (which do not have any primary keys and no foreign or
unique keys pointing to them).

During this process, A0 tries to automatically translate non-standard SQL constructs into forms
that fully conform to the standard, provided that this conversion will not lead to any loss of
information in the primary data. Otherwise the conversion is not performed. (The exact details of
this criteria are described in the SIARD documentation.) Correspondingly, objects that do not
conform to the SQL standard, nor can be automatically mapped to standard forms, are automatically
set to the status ``Cannot be archived''.

When finished, the results are presented to the user as shown in Figure \ref{FigA0}. There is a
hierarchical, collapsible tree of database objects in the left-hand pane, while the right-hand pane
shows the technical metadata from the data dictionary as well as the results of the analysis for
the database object that is currently selected in the object tree. The tree of database objects
comprises schemata, tables, table columns, one- and multiple-column primary key constraints, check
constraints, triggers, views, the views' SQL code, view columns, stored procedure, functions,
users, user roles, user role privileges, synonyms, and database links. There is either a small
colored bullet or a document symbol attached to each object in the tree. Additionally, each root of
a branch of the tree has assigned either a green check mark, a yellow triangle with an exclamation
mark, or a red cross. The colors and symbols indicate the status of an object:

\begin{itemize}
\item A green bullet means ``The object is or was made fully conforming with standard SQL and has
proper data integrity -- Ready for archiving''.
\item An orange bullet means ``The object is or was made fully conforming with standard SQL
but has problems with data integrity -- Ready for archiving''.
\item A red bullet means ``The object does not conform to and could not made conforming
with standard SQL -- Archiving is not possible without user intervention which may cause loss of
information''. Red bullets usually occur for unknown, proprietary or non-standard user-defined data
types.
\item A gray document symbol means ``The object does not conform to and could not made conforming
to standard SQL -- A0 decided to exclude it from archiving.'' This usually occurs when the SQL code
of a view, constraint, trigger, stored routine, or function does not conform to standard SQL.
\item A green check mark indicates that everything is okay (green bullets) or consolidated
(gray document symbols) on subsequent nodes of the respective branch. This branch is ready for
archiving.
\item An exclamation mark indicates that there is at least one warning (i.e. a orange bullet) on subsequent
nodes. The branch can be archived, though.
\item A red cross indicates that there is at least one unresolvable problem (i.e. a red bullet) on
subsequent nodes. A decision by the user is required (for example manual exclusion of the object
from archiving).
\end{itemize}

If an object was automatically excluded or has a red bullet, the detected problems are explained in
the``Details'' tab of the right-hand pane.

\subsubsection{\label{a0-clearance}Clearance}

Archiving of the database is not possible as long as red crosses appear (i.e. any red bullets on
lower levels of a branch), and the ``Create Archive" button is disabled. Thus, the minimal action
required by the user is to treat at least all objects with red bullets. To do so, the user has four
possibilities. Three of them can be chosen either from the function button panel or the context
menu of the right mouse button: First, the object may simply be manually excluded from archiving.
In this case, the red bullet turns into a gray document symbol. However, the red bullet may come
from an unknown data type which in fact is a valid standard SQL type but has a non-standard name
(thus SIARD does not recognize the type). This usually should happen only in the generic access
mode, but could also occur in expert modes when the version of A0's expert mode is older than the
version of the RDBMS. (In fact, many RDBMS use non-standard names for standard data types.)
Pressing A0's ``Unknown and Proprietary Data Type'' button, the user can define a catalog of
synonyms for data type names to disclose to A0 the correct data type for this ``unknown'' data type
name.

In the example shown in Figure \ref{FigA0}, all red crosses are caused by a single problem which
could not be resolved by A0's initial automatic conversion algorithm: The data type ``MY\_TYPE'' of
the column ``CODE'' in table ``PRINT\_MEDIA'' is an unknown data type and thus could not be
automatically converted to a standard SQL data type without loss of information. In consequence, A0
assigned it a red bullet, and the analysis result (visible in the right-hand pane) says ``SQL3
Check: Type conversion to SQL3 impossible''. A0 did not assign a gray document, however, since
there may be still the chance to solve a possible naming conflict by a user-supplied synonym rule
as described above. In this example, the data type ``MY\_TYPE'' is a user-defined data type with a
non-standard definition. However, inspection of the original database would reveal that (in this
example) ``MY\_TYPE'' can be mapped onto the standard SQL data type ``varchar(10)'' without loss of
information. The user can easily resolve the problem by defining a naming rule ``MY\_TYPE
$\rightarrow$ VARCHAR(10)'' using A0's ``Unknown and Proprietary Data Type'' button.

The second possibility is that the red bullet in fact comes from a proprietary non-standard data
type. The user may either exclude the object from archiving (third possibility) or use the
"Proprietary Data Type" button to define a deep conversion to a standard data type. This may cause
loss of information, though, and is only possible in expert modes\footnote{This conversion
functionality is not yet implemented in the current release of SIARD.} (since the input and output
of a conversion must be well defined). The fourth possibility is to perform a conversion of unknown
or proprietary data types within the original database (i.e. outside of SIARD's A0 application),
for example by using vendor-supplied specific SQL CAST functions \footnote{A similar  procedure may
be applied to non-standard SQL code in views, triggers etc. However, we think that it is not
desirable to change a database too deeply for the purpose of archiving. Furthermore, many RDBMS use
a proprietary procedural programming language for stored procedures which cannot be easily mapped
onto the SQL/PSM language (cf. subsection \ref{SQLstandardized}).}.

All conversions initiated in A0 are \textit{never} performed in the original database but only
after the extraction of the database from the RDBMS. We also emphasize that exclusion of an object
from archiving does not mean that the object will be invisible in the final archive. In contrast,
all meta information about the object (available from the original database) will be documented in
the archive (including its non-standard SQL code). But it means that the object (for example a
table, table column, check constraint, stored routine, or trigger) cannot be actively restored
anymore in a RDBMS later.

Of course, the user may also exclude any valid object (with a green bullet) from archiving, for
example if archiving is not desired due to appraisal decisions. The excluded object will still be
fully documented in the SIARD archive (though its data content will be excluded). We emphasize
that automatic and manual exclusion of objects (for example entire tables or single table columns)
is only possible because A0 restores a proper data integrity after any exclusion. For example, when
a table is excluded from archiving, the user is warned if there are any primary keys in this table,
or if any foreign or unique keys of other tables are pointing to that table. If the user confirms
exclusion, A0 removes the key constraints from the referencing tables (without a cascading deletion
of the table rows) and automatically also excludes from archiving all views, triggers, check
constraints etc. which contain references to the excluded table. A similar procedure applies if
single table columns are excluded. Finally, we note that any exclusion operation can be reversed
again by using A0's `Undo' function.

Orange bullets primarily indicate isolated tables (with no key constraints in it and no foreign
keys pointing to it) and do not require user intervention. However, using the context menu of the
right-mouse button or the function pane, the user may define its own primary as well as foreign and
unique key constraints manually \footnote{In contrast to objects which belong to the original
database, user-added key and check constraints cannot be excluded from archiving but only deleted
completely (using the `Delete' function or button). The delete function is disabled for all other
objects.} on any table, assisted by an interactive panel. A similar function and panel is provided
to define user-added check constraints (using standard SQL code only). Both possibilities may be
optionally used if linkage information or check constraints for the table do implicitly exist but
are hidden in external application software which operates on the original database. Adding the
corresponding information from the external application's system documentation will complement and
improve the database archive while external software usually will not be archived. (Note that
user-added key and check constraints are only defined within the SIARD archive, while the original
database is not altered.)

During work, the user can save intermediate states at any time, and resume his work later.
Furthermore, the application A0 has a special added-on object at the end of the object tree, named
`changelog'. In fact, A0 traces and remembers all changes to the original state of the database.
This includes activities performed automatically by A0 during its initial analysis of the original
database, as well as all changes caused by manual operations of the user later on. Each entry in
the `changelog' contains a time stamp, a short description of the activity, and the nature of the
change. The log file will be part of the SIARD archive (and be extended further in the subsequent
module A1 of SIARD). This enables re-tracing of the archiving process at any time in the future and
thus supports authenticity of the database archive.

\subsubsection{\label{a0-creation}Creation of the Archive}

The ``Create Archive'' button is enabled when the database is ready for archiving. This happens as
soon as all root nodes of branches in the object tree have assigned either green checkmarks or
yellow exclamation marks. The latter indicate warnings (orange bullets) on subsequent nodes (for
example isolated tables) which may be acceptable. For the example shown in Figure \ref{FigA0} this
state can be reached by either defining the before-mentioned synonym rule for the data type
``MY\_TPE'', or by exclusion of the table column ``CODE'' from archiving (while its metadata will
still be included in the SIARD archive).

When pushing the ``Create Archive'' button, A0 asks for the location where the archive shall be
saved. Afterwards, A0 starts to load the primary table data from the original database, performs
all necessary conversion operations on them\footnote{Depending on the amount of data and the
network transfer capacity, this may require a few seconds up to many hours. Primary data is loaded
and processed sequentially and will not necessarily have to be stored on A0's local machine. In
fact, the RDBMS, the application SIARD A0, and the SIARD archive files may all be at different
remote locations and connected by a network.}, and creates all SIARD archive files as depicted in
Figure \ref{FigWorkflow}:

\begin{itemize}

\item \textbf{SQL-DDL files}: These files contain standard SQL:1999 Data Definition Language (DDL) statements
only. Together they represent the definition of a self-contained relational database, comprising
all objects and attributes of the original database, except for those that were excluded.

\item \textbf{Table data files}: These are files which contain the primary data (except for
large objects, see below) of the database defined in the DDL files above. There is one file per
table. The data of one table row is contained in one line of the file, and data items have variable
lengthes. Rather than putting absolute delimiters between two adjacent data items, we use a simple
algorithmic token for delimitation\footnote{Every item has the form ``$l,\cdots;$'', where $l$ is
the number of characters of the data item ``$\cdots$''. (The semicolon is not required but used for
convenient reading by humans.) This representation is foolproof (whereas absolute delimiters are
not) and independent of the character encoding (whereas the byte length of the data item strongly
depends on it, in particular for variable-width multi-byte encodings like, for example, UTF-8).
This approach requires that every line is processed sequentially from the start, which is not a
disadvantage.}.

\item \textbf{Large Object Files}: Each file contains a single data item of a so-called large object
string type (provided that there really exist such data types in the original database, of course).
This is either a character large object\footnote{Actually, before writing the CLOB files, SIARD A0
converts the CLOBs of the original database into National Character Large Objects (NCLOB) via
translation of all characters (which are stored in the implementation-defined character encoding of
the RDBMS) to the fixed-length Unicode character encoding UTF-16. Otherwise, the characters in CLOB
files would be rather useless without exact knowledge of the original, implementation-dependent
character encoding.} (CLOB) or a binary large object (BLOB) which was embedded in a table row of
the original database. The former type is an arbitrary sequence of characters (for example a
narrative document encoded in XML), the latter is an arbitrary sequence of bytes (for example an
image file). Both may be up to several gigabytes in size. BLOB files just contain a hexadecimal
dump of the original BLOB, thus may contain anything\footnote{The ``expert modes'' of SIARD A0 also
use CLOBs and BLOBs to accommodate certain kinds of proprietarily typed large objects without loss
of information. For example, the expert mode for Oracle puts an Oracle BFILE (i.e. a BLOB that is
stored outside of the RDBMS) and Oracle LONG RAW objects into BLOBs, wile the expert mode for
Microsoft SQL-Server puts Microsoft TEXT and IMAGE objects into NCLOBs and BLOBs, respectively. }.

\item \textbf{XML Reference File} This XML document contains all information from A0's workbench.
More precisely, it contains three kinds of metadata: Firstly, the complete database logic that is
also contained in the SQL-DDL files (but encoded in XML). Secondly, all metadata from those
database objects that were excluded from archiving (and thus do not appear in the DLL files),
including the code of stored routines, triggers, views etc. Thirdly, the data from the
``changelog'' which reveals when and what actions where performed during the archiving process,
either automatically be A0 itself or manually by the user.

\end{itemize}

Note that a SIARD archive is made up of all files listed above, not only of the DDL files (which
only contains those parts of the original database which may be reloaded into another RDBMS again).
In particular, the XML Reference file accomplishes the smooth integration of archived database
objects, excluded objects, and all kinds of metadata. All files are plain text files (but may
contain hexadecimal strings for binary data), and Unicode/UTF-16 encoding
\cite{UnicodeSpec,ISO-10646-UCS,UTFspec} is used to overcome implementation-defined character
encodings of the original RDBMS, and to preserve multilingual character sets (including non-latin
alphabets).

The XML reference file may also be exploited for various other tasks. For example, it may be used
by XML schema mapping tools to generate metadata subsets for import into finding aid or catalog
systems. Or it may be used in the future for easy migration of the SIARD archive to other formats,
for example to a future release of SQL (though SQL is developed upward compatible).

Furthermore, for more convenient handling by humans only, every table data file has a short XML
header that contains the main table and column metadata such as names, data types, sizes, key
constraints (primary, foreign, and unique), or default values. These table data files will not be
altered anymore during the next steps of the SIARD workflow and therefore may be right away
utilized for other purposes. For example, they can be viewed with any trivial text editor program.
In addition, every SIARD archive contains a XML stylesheet language (XSL) file named
``dmpFile.xsl'' in its data file directory. Therefore, if a table data file is opened with a web
browser\footnote{This requires a browser which supports XSL, for example current releases of
Mozilla \cite{Mozilla} or Microsoft Internet Explorer} from within this directory, the user
automatically gets a pretty-print version of the data file (rather than clumsy to read raw data
rows), including named columns as well as vertical and horizontal table lines. For example, the
user may print this version (or catch the HTML output) for the purpose of non-technical
distribution of the data.

The SIARD archive has a fixed structure of file directories. This directory structure as well as
some other information is contained in an XML file named ``archiveInfo.xml'' that is located in the
root directory of each archive (and may be used by other software programs).

\subsection{\label{A1}A1: Description}

\begin{figure*}
\includegraphics[width=179mm,keepaspectratio]{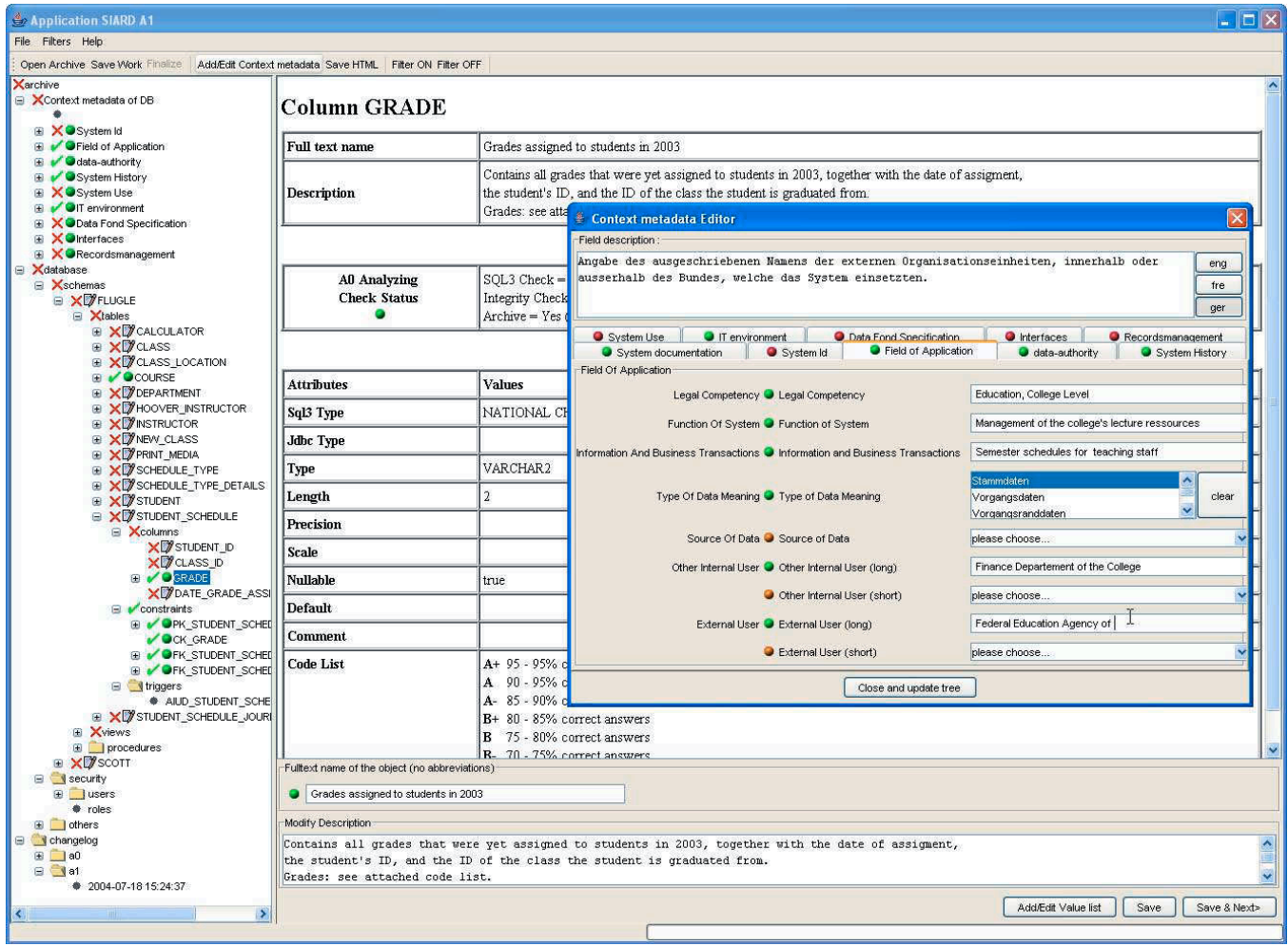}
\caption{\label{FigA1}A1's workbench to add complementary narrative and context metadata to the
database archive. For a single database object (selected in the object tree on the left-hand side)
metadata is entered in the right-hand pane, while context metadata on the database level is entered
in a multi-tab ``Context Metadata Editor'' window (shown in front of the right-hand pane).}
\end{figure*}

After creation of the SIARD database archive, the next step shown in Figure \ref{FigWorkflow} is to
add complementary metadata which is not available from the original database or the RDBMS. As
discussed in Section \ref{Intellectual}, this kind of metadata is indispensable to enable long-term
intelligibility and comprehensibility of the archived database.

Such metadata is added to the database archive using application SIARD A1. The users of A1 will not
have to be the same as those of A0. Instead, it may be more non-technical staff as, for example,
records managers, data asset managers, experimental scientists, or application responsibles. A1
only reads and modifies the XML reference file created by the application A0 (cf. Section
\ref{A1}), and the state of work my be saved at any time and resumed later. Thus the complementary
metadata for A1 may be gathered across different business units by simply forwarding this XML file
from one person to the other.

Figure \ref{FigA1} shows the graphical user interface of A1. Again, there is a database object tree
on the left-hand pane. It is basically the same as in A0, except for the colors which now have a
different meaning: red crosses indicate that there are mandatory but not yet filled in metadata
fields on subsequent nodes, while green checkmarks indicate that all mandatory metadata has been
provided on subsequent nodes.

For a database object that is selected in the object tree, the right-hand pane shows the metadata
that was contributed by A0 (and thus comes from the original database), plus metadata that was
added using A1. On the object level, the user can enter metadata at three locations in the
right-hand pane: An arbitrary full text object name which may be more meaningful than its technical
name (which is often an abbreviation), an arbitrary narrative description of the object which may
improve the intelligibility of the object, and finally a user-added code list. The latter may be
essential since code lists are frequently documented only outside a database. In our example in
Figure \ref{FigA1}, the column GRADE of the table STUDENT\_SCHEDULE contains grades that were
assigned to students after visiting certain college courses\cite{FLUGLE}. The grades are coded with
one or two characters (e.g. A+ or F-). However, the original database does not contain any
information about what these codes mean. Therefore, the user has added a code list (using the
``Add/Edit Value List'' button) which explains that, for example, A- means ``85 - 90\% of the exam
questions were answered correctly by the student''. Often codes are even less self-explanatory
(imagine a code ``92'' which in fact means ``application rejected'').

The context metadata on the database level is added and edited in a separate, multi-tab ``Context
Metadata Editor'' panel (see Figure \ref{FigA1}). It contains various tabs, each of them covering a
certain subject in the context of the database's usage, creation, original IT environment, history
of the system, data authority, or provenance. Each tab contains a number of metadata fields with
colored buttons that indicate whether filling in is mandatory or optional. A field is either a free
text field or a pull-down list to select from predefined values. If the user points into a field, a
description of the field is shown at the top of the panel. In addition, the descriptions can be
displayed in different languages (which can be selected at the upper-right corner of the panel).

The extend and structure of the context metadata usually depends on specific requirements of the
archival institution. Therefore, the ``Context Metadata Editor'' (CME) panel is fully customizable
by using SIARD's ``Context Metadata Schema Editor'' (CMSE) which is part of A1. Actually, the CME
is built from an XML file which contains the CME layout, and the CMSE is basically a graphical user
interface to manipulate this XML file. The user may define its own CME panel, including individual
tabs as well as individual metadata fields and descriptions (using arbitrary languages). Thus it is
possible to define several standardized context metadata schemata (CME panels) for different
classes of databases or archival scopes. Depending on the class or purpose, the appropriate CME to
be loaded in A1 to add customized context metadata to the SIARD archive.

In addition, there is a special CME tab ``System Documentation'' which allows the user to enclose
arbitrary files to the context metadata part of a SIARD archive\footnote{Note that A1 does not
inspect these files at all. It's up to the user's responsibility to use document and data formats
that are suited for long-term preservation.}. These may be, for example, PDF documents or TIFF
images that are taken from the original RDBMS and database documentation, for example system and
user manuals, log files, security reports, original data dictionaries etc. (One could even think of
MPEG video documents showing people at work with the original production database system.)

While the user is working with A1, the application records all user modifications in the subnode
``A1'' of the``changelog'' which was already mentioned in Section \ref{a0-creation}. Furthermore,
for all context metadata entered through the CME panel, A1 does not only store the filled in data
values of a metadata field but also the entire, language-specific version of the field description
which was visible in the panel at that moment when the user entered or modified the data value.
Making the field descriptions and the mandatory/optional characteristic integral parts of the SIARD
archive is necessary since metadata catalogs (i.e. CME panel configurations) may change or evolve
over time, and the meaning of single metadata items may change.

At any time, the user may convert the current version of the XML reference file into an HTML
document for convenient review or distribution. When all mandatory metadata has been filled in, the
``Finalize'' button is enabled. When this button is pushed, A1 will warn the user that no further
changes to the reference file will be possible later, and will then write the final version of the
XML reference file. This version now contains all metadata that was generated or collected by A0
and A1. The SIARD database archive has been completed and is now suitable long-term preservation.
It does not depend on any specific hardware and software (not on SIARD as well), it consists of
plain text files only (except for optional, user-added PDF or TIFF files), and it is solely based
on technologies which are widely used and internationally standardized.

\subsection{\label{A2}A2: Reload and Access}

\begin{figure*}
\includegraphics[width=179mm,keepaspectratio]{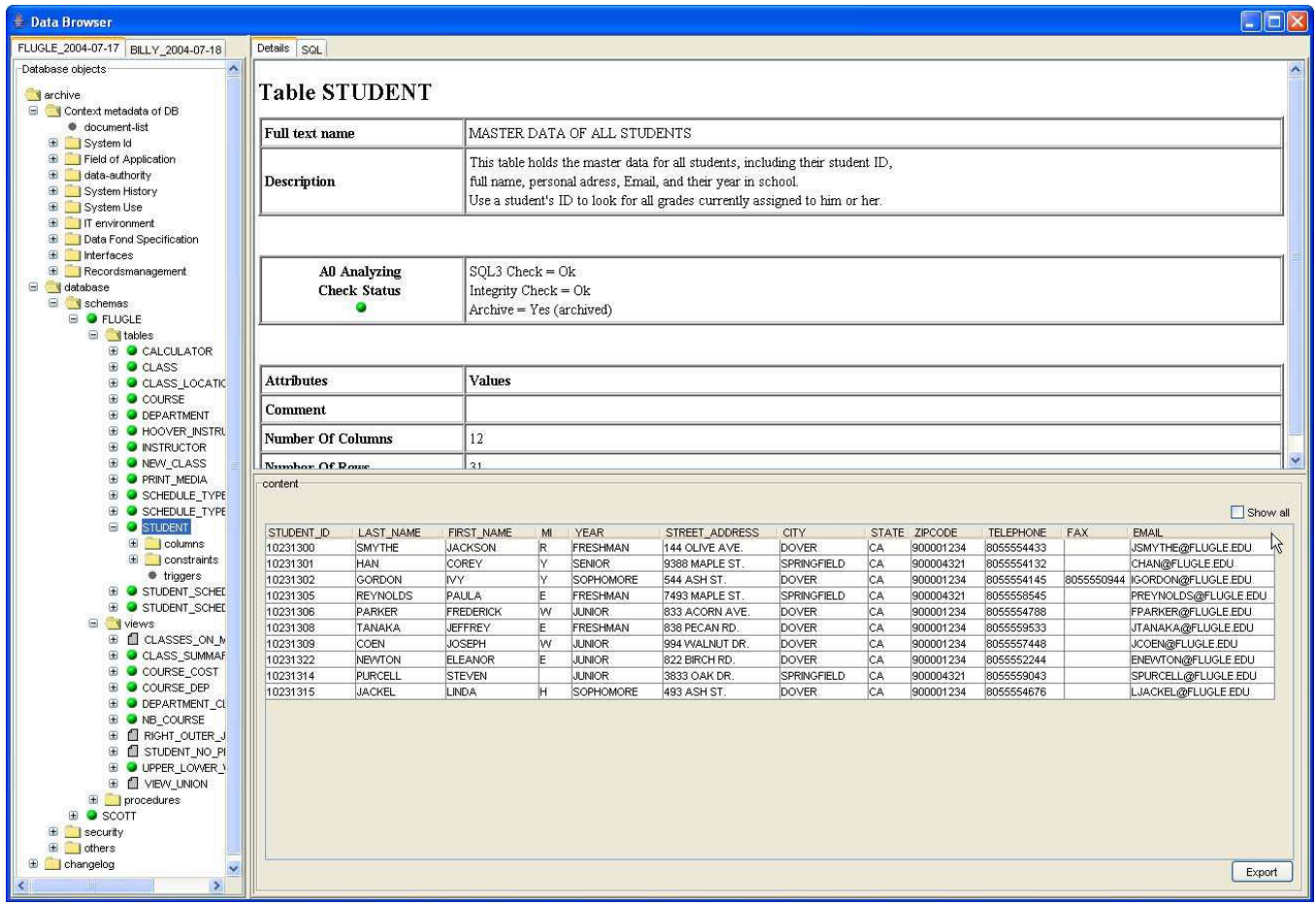}
\caption{\label{FigA2}A2's Data Browser after reload of two SIARD database archives into an Oracle
database. The pane at the bottom shows the data of the table which is selected in the database
object tree (left-hand pane), while the pane at the top shows its metadata. The other database
(``BILLY'') is accessed by selecting the second tab of the left-hand pane.}
\end{figure*}

Apart from their primary purpose, SIARD database archives may also be utilized and processed by
other software applications, for example in data dissemination and exchange or data warehouses.
Because of its open technologies and its high level of standardization, SIARD archives may also be
easily converted into forms that can be accessed directly in the World Wide Web. However, this may
require some additional, expensive hardware and complex software (e.g. an application web server).

We have therefore added a third application to SIARD, named A2, which does not require additional
infrastructure at all. The same infrastructure that was used for archiving the database (i.e.
creating a SIARD archive) will be perfect. Basically, A2 is a simple RDBMS client but enables users
to reload SIARD archives into a RDBMS ``on demand'', and then provides multi-user remote network
access for querying and browsing the data together with its technical and descriptive metadata in
one graphical user interface.

We emphasize that A2, in its current version, is only a prototype (thus still has some bugs,
limited functionality, and some security deficits), and currently only works with an Oracle RDBMS
as the reload target system. However, it will be rather simple to adapt A2 to work with other RDBMS
products as well.

Working with A2 requires a running standard Oracle database instance (reachable either through a
TCP/IP network or the local host) and some very simple preparations prior to operate A2. This
preparation, however, has to be done by a system administrator:  A new tablespace to hold the
restored SIARD schemata and data should be created (and provide enough space for the estimated
SIARD restore operations). In addition, a user with specific rights (which are described in the
user manual) has to be added to the RDBMS. This user will act as a ``SIARD archive manager'' to
control and serve connections from A2 clients. Its database schema will be the repository for
information about all reloaded archives, registered users that are allowed to reload SIARD archives
to this RDBMS, and it records reload operations and connection times of individual users. This
schema will be initialized automatically as soon as any SIARD A2 client is connecting to this RDBMS
for the first time.

A2 is controlled by an XML configuration file which contains the profiles of one or more RDBMS that
are available as targets for SIARD restores. A profile identifies the Oracle database instance, the
tablespace used for restore operations, and the before-mentioned ``SIARD archive manager'' user.
Moreover, it provides additional information required by A2 to connect to the RDBMS. This
configuration file will be pre-configured and provided to end-users of A2 by the RDBMS system
administrator.

After starting A2, the user is asked to choose (from a file selector panel) the ``archiveInfo.xml''
file from the SIARD database archive to be reloaded, to select one of the pre-configured connection
profiles, and to decide whether or not large object string files (BLOB and NCLOB, cf. Section
\ref{a0-creation}) shall be reloaded too. Finally, the user either pushes the ''Create New User''
button to create a personal database account for working with A2, or else chooses an existing
account from a pull-down list. After entering the user password\footnote{Warning: This user
password only authenticates the human user of A2, not the connection between A2 and the RDBMS
itself, whereas A2 connects to the RDBMS as the ``archive manager'' user which has granted DBA
rights, and its password is contained as clear text in the A2 configuration file! This is a severe
security risk. As mentioned: A2 is still a prototype, and a production release of A2 will have to
contain a secure authentication of A2 clients. Thus we \textit{strongly} recommend \textit{not} to
use A2 with a production database (i.e. one which holds any important data aside from SIARD
database reloads).}, A2 reloads all schemata contained in the SIARD database
archive\footnote{Depending on the size of the archive and the network transfer capacity, this may
require a few seconds up to many hours.}. The user may sequentially reload as many SIARD database
archives as needed and use them all together in a single A2 session. If a SIARD database archive
was already reloaded by another user, it will not be reloaded a second time. (Users of A2 only have
read access to the reloaded database.)

Figure \ref{FigA2} shows the main component of A2, the ``Data Browser'' panel. Again, there is the
database object tree navigator already known from A0 and A1. In addition, there are tabs at the top
of the tree panel to switch between different reloaded databases. The pane at the bottom shows the
data of the table which is selected in the database object tree, while the pane at the top shows
its metadata. Further metadata is found on subsequent nodes and in the ``Context Metadata" branches
below the ``archives'' root node. For example, the code list for student grades (which was
user-defined during work with A1 in Section \ref{A1}) can be accessed on the corresponding leave
node ``code list'' in the FLUGLE schema (table STUDENT\_SCHEDULE, column GRADE).

As can be seen in Figure \ref{FigA2}, there are some (but not all) views reloaded. These views
(marked with a green bullet) are actually proper views, and selecting them will show the view's
data in the data pane. The other views (those with a grey document symbol) do not show any data
because they were excluded from archiving by SIARD's application A0 due to non-standard SQL code in
the original database. However, the view's definition as well as the reason why it was excluded
will be visible in the upper metadata pane.

The user may browse the database, select parts of table data in the data pane, and export it as
comma separated (CSV) files. Furthermore, users who are familiar with the SQL query language may
use the ``SQL'' tab (at the top of the upper pane) to switch to a separate panel where arbitrary
SQL:1999 query statements can be composed and send to the Oracle RDBMS. The results of the query
will be shown in a second pane where they can be selected all or in parts for export to a CSV text
file.

If a user A decides to quit A2, he or she will be asked if the restored database shall be deleted.
However, if there is still another user B currently registered for using the same database, it will
not be deleted (but the registration of user A for this database will be removed).

In conclusion, SIARD A2 seamlessly integrates all metadata and data (whereas the reloaded database
only contains the metadata from the DDL files described in Section \ref{a0-creation}), and it
enables the user to perform simple as well as complex SQL queries on restored databases. Query
results can be exported to the local machine. Several A2 clients may connect simultaneously to the
same restored database, and A2 provides controlled multi-user remote access\footnote{Any Firewalls
between the A2 client and the RDBMS will have to be properly configured, though.} to SIARD database
archives.

\subsection{\label{Development}Development Environment}

Software development of SIARD was carried out by the Swiss Federal Archives together with Trivadis
(Switzerland) AG \cite{Trivadis}. The SQL parser and validator as well as the context metadata
schema editor was developed by one of the authors (SH).

The SIARD software is platform independent, relying on the programming language Java and the Java
virtual machine as an interface to the operating system. It was tested under Solaris 7 and 8, Red
Hat Linux 7, and Windows NT / 2000 / XP. The JDBC driver may need a specific environment to
run properly (e.g. an MS Windows for a Windows specific authentication on an MS SQL-Server), but these
restrictions are solemnly defined by the driver.

Currently, Eclipse \cite{Eclipse} is used as Integrated Development Environment, mainly because
of its file-based approach. The technologies used in SIARD are

\begin{itemize}
\item Java 2 Software Development Kit (J2SDK) 1.4 by Sun Microsystems \cite{Java}.

\item Java Foundation Classes / Swing \cite{Java} were used for the graphical user interface.

\item Java Database connectivity (JDBC) 3.0 \cite{Java} for data retrieval from and reload to the
database management systems.

\item Java API for XML Processing (JAXP) 1.2 \cite{Java} which handles operations on XML data (not needed
anymore for J2SDK 1.4.2 and higher).

\item The Extensible Markup Language (XML) 1.0 \cite{XMLspec} for semantic markup of data in
external text files created and read by SIARD components.

\item The Extensible Stylesheet Language Family (XSL): XSL Transformations (XSLT) Version 1.0
\cite{XSLTSpec} is a XML-based declaration language for displaying and transforming XML files.

\item XML Schema (XSchema) 1.0 \cite{XMLSchemaSpec} provides automatic consistency and
integrity checks on the SIARD XML files.

\item The Structured Query Language (SQL) ISO/IEC 9075:1999 \cite{ISO-9075-SQL1999} for definition
of database layouts.


\item The Unicode Transformation Format UTF-16-UCS-2 \cite{UnicodeSpec,ISO-10646-UCS,UTFspec} for
platform independent and multilingual character encoding in all SIARD text files.

\item Oracle 8/9i and Microsoft SQL-Server 7/2000 relational database management systems and Microsoft
Access 97/2000 were used for testing SIARD.

\end{itemize}

The JDBC driver implementations of the different database manufacturers provide a varying degree of
compliance with the JDBC specifications. Especially, the functions for querying the database for
metadata leave much to be desired. This metadata could be extracted from the database by database
specific SQL-like queries, which is why the database access in SIARD is encapsulated in so-called
``modes'', allowing for database-dependent enhancements.

``Expert modes'' allow manufacturer-specific access to the database engines. Today, expert modes
have been implemented and tested for JDBC Drivers for the Oracle and Microsoft products mentioned
above. For the use with other RDBMS products, a generic mode is provided. The open architecture of
SIARD allows the simple development of further expert modes for archiving from other database
products since they can be added dynamically (i.e. without changes in existing code).

As explained in Section \ref{a0-creation}, SIARD ultimately produces plain text files only. Testing
conformance of the SIARD XML files with the XML standard is simple: all have dedicated XML schemata
and thus can be validated, and there exist a multitude of XML parsers which provide just this
functionality. The compliance of the SQL structures files with standard SQL:1999 is governed by
SIARD's own, plugged-in SQL parser and validator that checks the syntax of the SQL expressions as
well as the dependencies between these expressions. (For example, for a table to be created in a
specific database schema, the schema must have been created first.) Independent cross-checking of
SIARD's SQL files is possible too, though we are aware of only one other tool which provides broad
SQL validation functionality \cite{MimerVal,SQuirall}.

\section{\label{Conclusions}Conclusions}

We have discussed problems and relevance of long-term preservation of relational databases for
usual archival institutions like national archives and scientific data archives that have to ingest
data from a broad diversity of vendor products. We have argued that the common current ingestion
and preservation practices may suffice for ingestion of small and simply structured data sets but
suffer from insufficient integration of data and metadata, lack of automation in the ingestion of
large amounts of data, error-proneness, and in general require extensive manual effort and
intervention to make the data accessible and usable. Without having more efficient solutions at
hand, the rapidly growing size and complexity of relational databases in modern relational database
management systems will rapidly outpace the ability of archives to ingest, manage, and preserve
them.

Furthermore, we have surveyed the relation between present-day database management system products
and the standardized data definition, query and manipulation language SQL, and have critically
appreciated its applicability to long-term preservation of databases. From this discussion we
concluded that widespread non-standard, vendor-supplied enhancements and additions do not allow for
one-to-one ingestion from such systems, and that no "SQL for Archiving" exists. Nevertheless,
standard SQL may be reasonably exploited for long-term preservation purposes when data and data
logic are actively extracted from database management systems by specialized ingest tools which map
different "SQL flavors" to generic SQL, and transparently trace and document those parts which
cannot be mapped.

Finally, we have presented the method and platform independent application ``Software-Invariant
Archiving of Relational Databases''(SIARD), developed at the Swiss Federal Archives. It is a an
efficient, traceable and controllable ingest tool to detach relational data from any specific
hardware and software environment, and thereby enables, up to a reasonable level, retention of its
original authenticity, integrity, accessibility, and usability. SIARD integrates data with data
logic and descriptive metadata and supports the intelligibility of databases for long-term archival
preservation and access.

\vspace{3mm} \textsf{\scriptsize Readers who are interested in testing and reviewing SIARD or
writing new expert modes may contact the authors. SIARD is property of the Swiss Federal
Administration.}



%

\end{document}